\documentclass[aps,prl,floatfix,showpacs,preprintnumbers,amsmath,amsymb,a4paper,superscriptaddress,twocolumn,longbibliography]{revtex4-1}
\usepackage{amsmath}
\usepackage{graphicx}
\usepackage{amssymb}
\usepackage{amsmath}
\usepackage{epstopdf}
\usepackage{float}
\usepackage{color}

\newcommand{ \bi}{\begin{itemize}}
\newcommand{\ei}{\end{itemize}}
\newcommand{\be}{\begin{equation}}
\newcommand{\ee}{\end{equation}}

\begin{document}


\title{Extreme correlation and repulsive interactions in highly excited atomic alkali anions}


\author{Matthew T. Eiles}

\affiliation{Department of Physics and Astronomy, 
Purdue University, 47907 West Lafayette, IN, USA}
\author{Chris H. Greene}

\affiliation{Department of Physics and Astronomy, 
Purdue University, 47907 West Lafayette, IN, USA}
\affiliation{ Purdue Quantum Center, Purdue University, West Lafayette, Indiana 47907, USA.}


\date{\today}

\begin{abstract}
At high energies, single-photon photodetachment of alkali negative ions populates final states where both the ejected electron and the residual valence electron possess high angular momenta. The photodetached electron interacts strongly with the anisotropic core, and thus the partial cross sections for these channels display non-Wigner threshold behavior reflecting these large, and occasionally repulsive, interactions. Our fully quantum mechanical theoretical study enables a deeper interpretation of these partial cross sections. Comparisons of the behavior in different channels and between different atomic species - sodium, potassium, and cesium - show the critical role of near-degeneracies in the energy spectrum and demonstrate that much of the behavior of the partial photodetachment cross sections stems from the permanent, rather than induced, electric dipole moments of these nearly-degenerate channels. This provides a concrete example of a system where negative dispersion forces play a decisive role.
\end{abstract}


\maketitle

Atomic negative ions are fertile sources of information about correlated electron behavior, shape and Fano-Feshbach resonances, and near-threshold behavior \cite{Andersen,PeggReview}. Negative ions of alkali atoms have an electron affinity around 0.5 eV \cite{HotopReview,Andersen2} and possess only one weakly bound state \cite{BuckmanClark}. At higher energies a rich spectrum of rapidly autodetaching doubly excited states appears \cite{LiuNa,LiuK,HanstorpK1,HanstorpLi,ExpLi1}. Much of the interest in negative ions stems from the fact that, unlike positive ions or neutral systems, they are bound together not by the Coulomb potential but instead by far weaker polarization potentials which reveal subtle correlation effects. Furthermore, the alkali anions focussed on here are effective two-electron systems and thus are theoretically tractable to high accuracy \cite{WatanabeGreene,Watanabe,OrangeRMP}. 

In the absence of dominant Coulomb forces, the structure of anions is determined by polarization potentials between the induced dipole moments of the extended electronic states and the additional electron \cite{WatanabeGreene}. These potentials cause the observed partial cross sections (PCSs) to deviate from the Wigner threshold law (TL), $\sigma\propto E_e^{l+1/2}$, where $E_e$ and $l$ are the photoelectron's energy and angular momentum \cite{Wigner1948}. This was first noticed in photodetachment experiments of alkali anions just above the first excited threshold, where the relevant ground state polarizabilities $\alpha_p$ are a few hundred atomic units and the Wigner TL fails surprisingly rapidly \cite{TaylorNorcross,LineBerger}; this sparked the development of several improved theoretical descriptions \cite{OMalley,Farley,Watanabe,RuzicMQDT,WatanabeGreene,ChrisJJ}. These polarizabilities increase rapidly with the principal quantum number $n$, approximately as $n^7$; for states with $n \approx 6$ and having large angular momenta, $l_{max} \approx n-1$,  $\alpha_p\approx10^4 - 10^6$ atomic units. At sufficiently high $n$ and maximal $l$, $\alpha_p$ can become negative, leading to an entirely repulsive potential \cite{HanstorpK2}.

These long-range induced dipole potentials typically dominate the low energy photodetachment spectrum of alkali anions. H$^-$ is exceptional owing to its ``accidental'' degeneracy. The degenerate states hybridize in the detached electron's electric field and form permanent dipole moments, characterized by a set of dipole parameters $a_i$ \cite{GailitusDamburg}. The resulting permanent dipole (hereafter called dipole) potentials differ remarkably from the induced dipole (hereafter called polarization) potentials, particularly if $a_i\le\frac{1}{4}$. In this case the potential supports an infinite number (which becomes finite since the degeneracy is always broken at some level) of doubly-excited states. Such sequences of resonances have been extensively studied theoretically and verified in an impressive series of experiments \cite{Sadeghpour1,Sadeghpour2,RostBriggs1990,RostETAL1991,LosAlamosH1,LosAlamosH2}.  Positive $a_i$ also exist, leading to repulsive potentials.  One compelling question is if this dipole structure is present in non-hydrogenic atoms, since the non-penetrating high-$l$ states rarely interact with the core and become nearly degenerate. Many parallels between hydrogen and lithium have been observed at higher ($n\ge4$) thresholds \cite{LindrothHLi,PanStaraceGreene1,PanStaraceGreene2}, but with other atoms largely unexplored.

Recently, the GUNILLA group at the University of Gothenburg measured PCSs for photodetachment into very excited channels: $7s$, $5f$, and $5g$ in potassium, $5g$ in sodium, and $10s$, $6f$, $6g$, and $6h$ in cesium  \cite{HanstorpK2,HanstorpPRA, HanstorpNa,HanstorpCsPRA}. These observations highlighted the dramatic role of the long-range interaction between the photodetached electron and the highly polarizable atom, especially in the unusual scenario involving repulsive interactions. The present Letter supports these observations with a fully quantum-mechanical calculation utilizing comparisons between atomic species and three theoretical probes - PCSs computed with the eigenchannel R-matrix method, a study of the adiabatic potential energy curves, and an analysis of threshold laws - to identify the essential physics and demonstrate that the hydrogen-like character of these highly excited states dominates, and thus the system is governed by permanent, rather than induced, dipole potentials. This improved physical model leads to a far more satisfactory interpretation of the observed threshold behavior, especially at higher energies where the polarization potentials lead to qualitatively incorrect predictions. To put this study in a more recent context, note that repulsive dispersion forces have been utilized or even designed to suppress undesirable and problematic inelastic collisions in quantum gases \cite{GorshkovZoller2008prl,MicheliZoller2007pra,deMirandaBohnETAL2011}.  Accordingly it is of interest to explore them in the present study's comparatively simple situation. 

The eigenchannel R-matrix method has experienced success in describing alkali atomic anions \cite{PanStaraceGreene2,LiuLi,LiuNa,LiuK}. Only a brief discussion is given here since Ref. \cite{OrangeRMP} and its references contain a detailed description. This calculation first determines the eigenspectrum of the neutral atom, confined to the R-matrix volume, using a B-spline basis to solve the one-electron Schrodinger equation with an $l$-dependent model potential \cite{LiuK,Marinescu}. The radius $r_0=250\text{a.u.}$ of the R-matrix volume encompasses the excited atomic states so that only one electron has non-vanishing probability outside the volume; it also is large enough to include additional channel coupling. 98 closed-type one-electron radial orbitals, vanishing at $r = 0$ and $r = r_0$, along with 2 open orbitals which do not vanish at $r = r_0$, are obtained for each partial wave $l=0-14$.

  \begin{table}[b]
\begin{center}
\setlength\tabcolsep{2 pt}
\begin{tabular}{ | c |c | c | c |}
\hline
\setlength\tabcolsep{2 pt}
  State & $\alpha(l,l_-), \alpha(l,l_+)$ &  State & $\alpha(l,l_-), \alpha(l,l_+)$\\
  \hline\hline
   Na$(3s) $ & 0.00, 166 &
    Na$(5d) $ & 4.93(6), 5.06(6)  \hphantom{1}\\Na$(5f)$ & 2.10(7), 2.12(7) &
     Na$(5g) $ & -2.55(7), -2.56(7)  \hphantom{1}\\
     K$(4s)$ &0.00, 308  & 
      K$(5f)$ & 5.01(6), 5.05(6) \hphantom{1} \\ K$(5g)$ &-5.14(6), -5.18(6) &
      K$(6f)$ & 2.65(7), 2.68(7)  \hphantom{1}\\ K$(6g)$ &4.46(7), 4.47(7) &    
      K$(6h)$ & -7.17(7), -7.19(7)  \hphantom{1}\\ 
    Cs$(6s)$&0.00, 445 &
    Cs$(6f)$ &7.57(6), 7.64(6)  \hphantom{1} \\ Cs$(6g)$ & 1.70(7), 1.70(7)&
    Cs$(6h)$ &-2.44(7), -2.45(7) \hphantom{1}\\
 \hline
\end{tabular}
\begin{tabular}{ |c|c|}
\hline
degenerate levels & $a_i=\lambda_i(\lambda_i+1)$ \\
\hline
\hline
$5f\epsilon l_\pm;5g\epsilon l_\pm$ & 47.9, 31.5,  2.09,-13.5 \\
$5d\epsilon l_\pm;5f\epsilon l_\pm;5g\epsilon l_\pm$ & 57.6, 43.0, 23.3, 8.32, -18.8, -31.4\\
$6f\epsilon l_\pm;6g\epsilon l_\pm;6h\epsilon l_\pm$ & 80.0, 61.6, 34.0, 15.0, -21.9, -38.6\\
\hline
\end{tabular}
\caption{Channel-dependent static polarizabilities $\alpha$ and dipole parameters $a_i$ in atomic units for $L=1$ and odd parity. As allowed by dipole selection rules, $l_-=l-1$, $l_+ = l+1$, and (A) represents $\times 10^A$. }
 \label{tab:polarizabilities}
 \end{center}
\end{table}

These one-electron functions form a two-electron basis which is used to variationally compute the eigenchannel representation of the R-matrix \cite{OrangeRMP,GreeneKimSL}. More than 18000 basis states in the final symmetry are used to ensure convergence for such a large $r_0$. Multipole interactions extend beyond $r_0$, so the coupled channel equations without exchange are propagated between $r_0 \le r \le  2000$a.u.  \cite{PanStaraceGreene2,ALLISON}.   These solutions, matched to the values at $r_0$ computed by the R-matrix, determine the K-matrix and dipole transition amplitudes, and therefore the cross sections. To understand the behavior of these PCSs it is advantageous to study the adiabatic potential energy curves, i.e. the eigenvalues of the potential matrix $V_{ij}(r)$ at fixed $r$ \cite{PanStaraceGreene2,WoodGreene}, as well as its asymptotic behavior:
\begin{equation}
\arraycolsep=-6pt\def\arraystretch{1.8}
V_{ij}(r) = \left\{\begin{array}{lr}\,\,\,\left(\frac{l(l+1)}{2r^2} - E_{i}\right)\delta_{ij}+\sum_{k=1}^\infty \frac{d_{ij}^k}{r^{k + 1}},&r> r_0\\
  \,\,\,\left(\frac{l(l+1)}{2r^2} -\frac{\alpha_p}{2r^4}- E_{i}\right)\delta_{ij},& r\to\infty\\
  \,\,\,\left(\frac{a_i}{2r^2} - E_{i}\right)\delta_{ij},\,\,\,\,\,\,\,\,\,\,\, r\to \infty,&\{E_i\}\to E_n.
  \end{array}\right.
\end{equation}
 $E_{i}$ is the atomic energy and $d_{ij}^k$ is a transition matrix element for the $k$th multipole moment. The sum is truncated at $k=3$. At large $r$ the $k=1$ term dominates and adiabatic diagonalization of $V_{ij}(r)$ yields the second expression, containing a channel-dependent polarization potential proportional to $\alpha_p$, the polarizability for the $i$th atomic state. Note that a negative polarizability gives a repulsive polarization potential. In the quasi-degenerate subspace of energies $\{E_i\}$ near a hydrogenic energy $E_n$, the third expression with a dipole potential becomes valid. Table \ref{tab:polarizabilities} gives relevant dipole parameters $a_i$. In the repulsive case $a_i>0$ and the dipole potential is a centrifugal potential with positive non-integral angular momenta $\lambda_i$; these become negative when $-\frac{1}{4} \le a_i \le 0$ and complex when $a_i<-\frac{1}{4}$:  $\lambda_i = -\frac{1}{2}+\sqrt{a_i+1/4}$.

\begin{figure}[t]
\begin{center}
\includegraphics[width =\columnwidth]{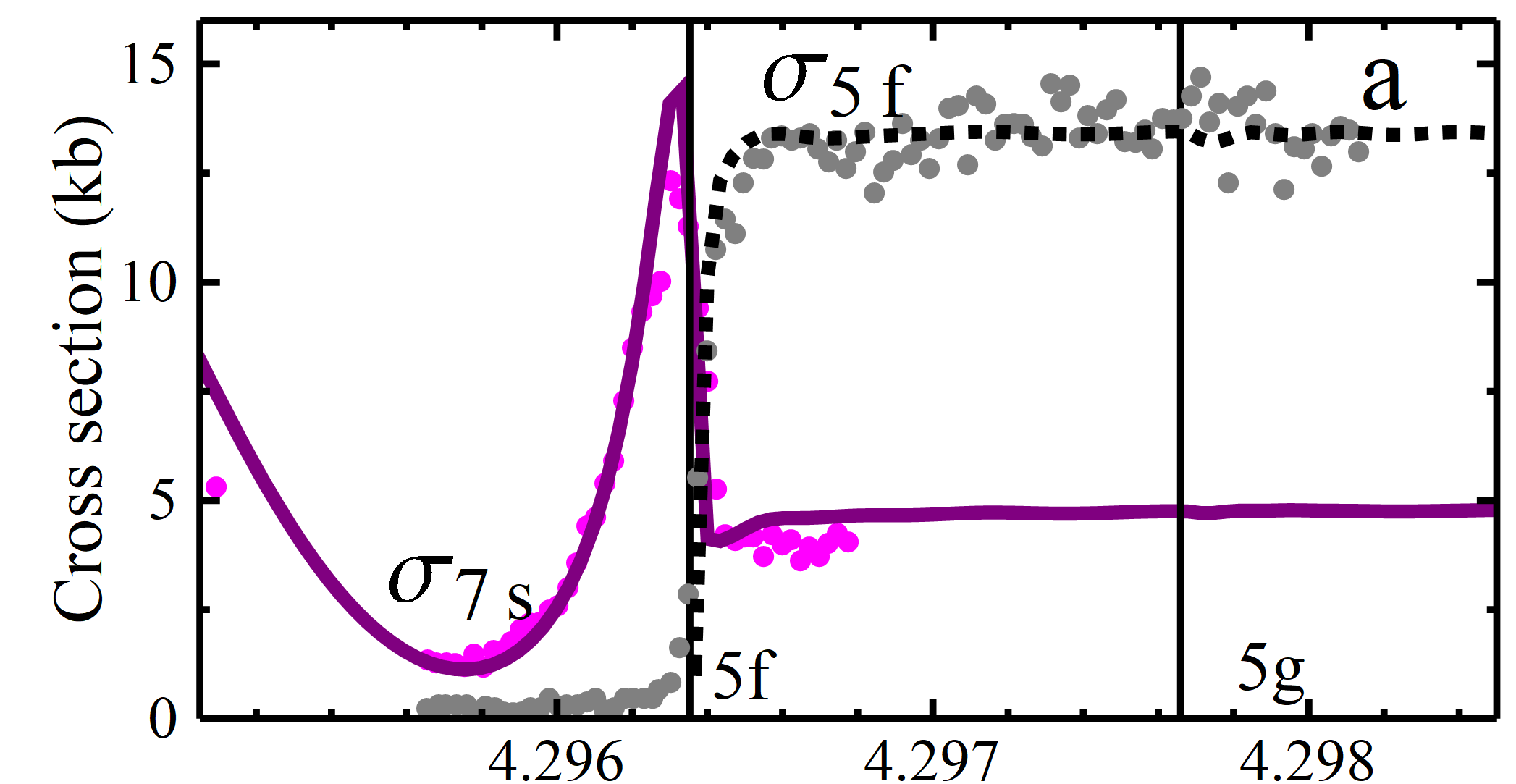}\\\includegraphics[width=\columnwidth]{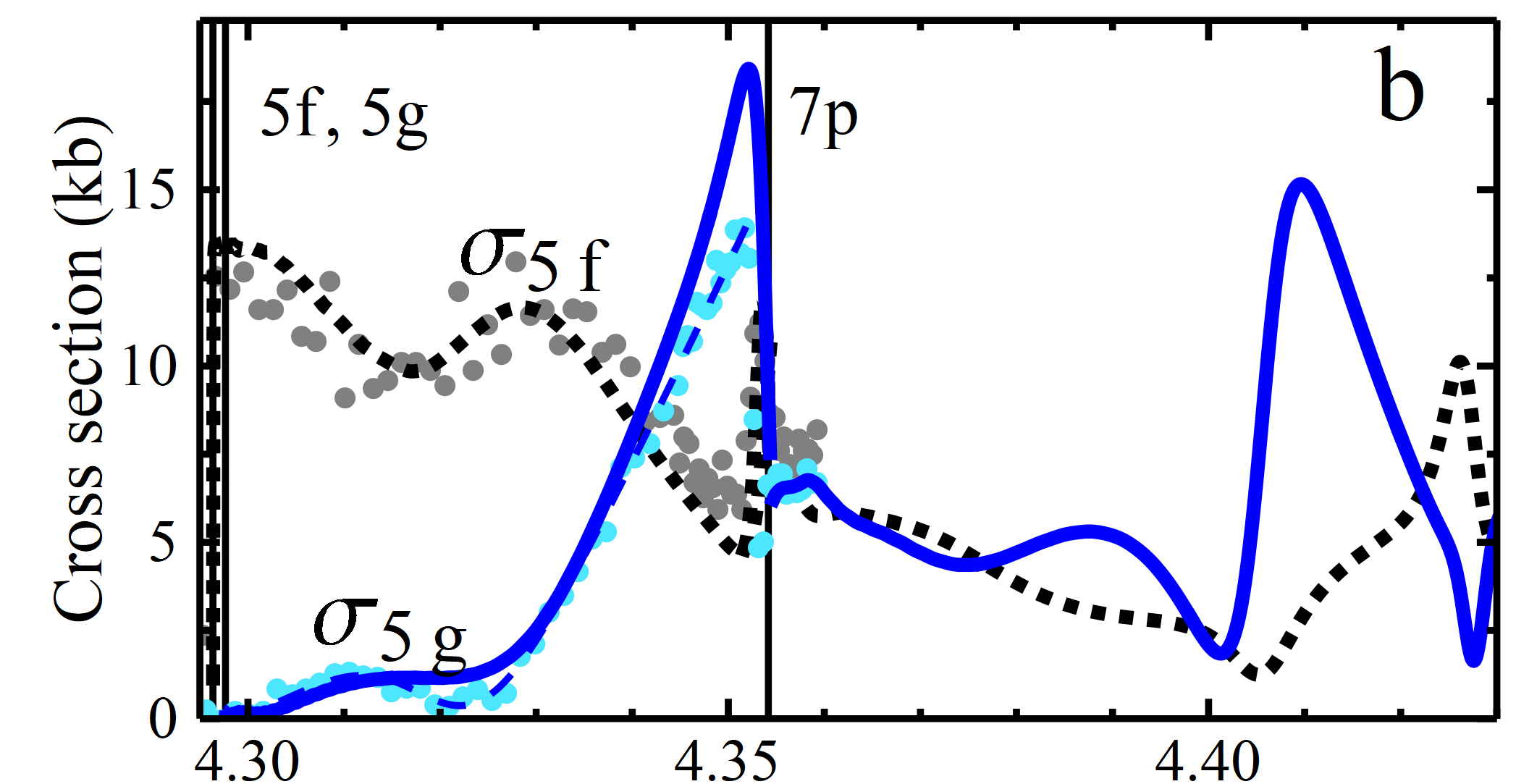}\\\includegraphics[width =\columnwidth]{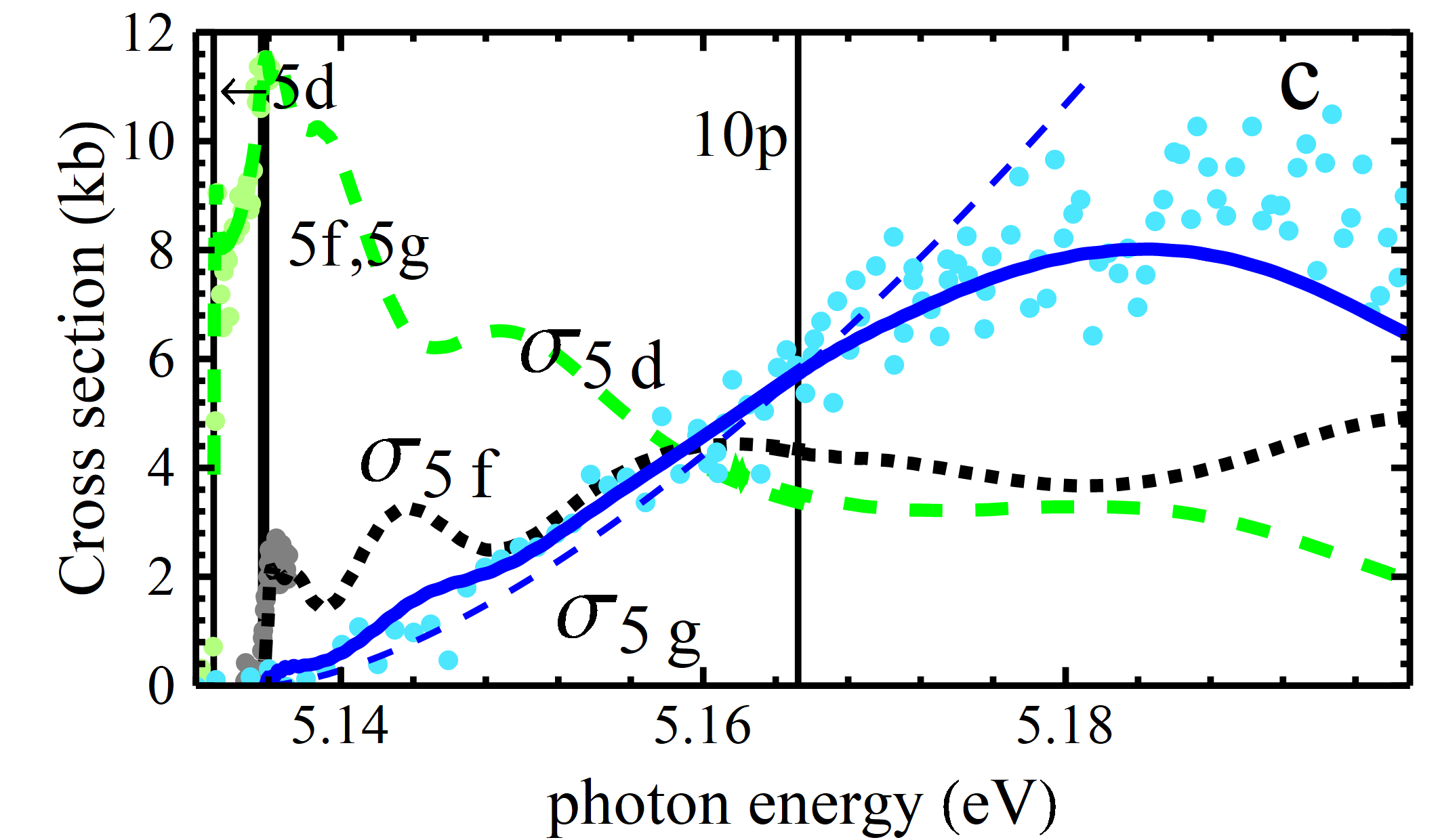}
\end{center}
\caption{\label{fig:5fzoom} Observed \cite{HanstorpK2,HanstorpPRA,HanstorpNa,RohlenThesis} (round dots) and calculated PCSs for a) potassium near the $5f,5g$ thresholds, b) potassium, c) sodium.  PCSs for the $7s$ (purple curves, panel a only), $5f$ (black, square dots), 5g (blue, solid), and 5d (dashed green, panel c only) are shown.  Thin dashed curves show TL fits.}
\end{figure}
 
These potentials define TLs which govern the PCSs.  Ref. \cite{HanstorpK2,HanstorpK3} developed semiclassical arguments for the threshold behavior for large polarizabilities. For attractive potentials, the polarization potential reduces the centrifugal barrier, so the photodetachment process rapidly saturates at energies exceeding this barrier, typically a few $\mu$eV. In the more unusual case of a repulsive polarization potential, Ref. \cite{HanstorpK2} developed an approximate TL using WKB arguments: $\sigma\sim \exp[2.850|\alpha_pE_e|^{1/4}]$. 
The final wave function must tunnel under the repulsive potential to overlap the initial state, so the transition dipole elements are small. In the case of approximately degenerate thresholds, repulsive potentials lead to a TL $\sigma\sim E_e^{\lambda_\text{min} + \frac{1}{2}}$, while in an attractive potential  the PCSs begin discontinuously at a finite threshold value \cite{GailitusDamburg}.

Fig. \ref{fig:5fzoom}a  shows calculated $7s$ and $5f$ PCSs, highlighting the accuracy of the R-matrix method by resolving the narrow resonance in the $7s$ channel and revealing the threshold behavior of the $5f$ PCS, which rises rapidly over a few $\mu$eV  before saturating, in excellent agreement with experiment \cite{HanstorpK2}. The induced and dipole TLs, both attractive in this channel, predict the same qualitative behavior. Fig. \ref{fig:5fzoom}b shows the $5f$ and $5g$ PCSs over the energy range studied in Refs. \cite{HanstorpK2,HanstorpPRA}. Here, the $5f$ and $5g$ thresholds are, to an excellent approximation, degenerate, and in accordance with the dipole TL the $5f$ PCS rises essentially discontinuously at threshold.  These calculations agree quite well with the experiment, and additionally the length (shown) and velocity gauge results are in excellent agreement. Total cross sections over this range agree with the calculation of Liu \cite{LiuK}, but he did not present results for PCSs in this range.  A time-delay analysis reveals a resonance at approximately 3.2 eV, in rough agreement with Liu \cite{LiuK} and experimental fits \cite{HanstorpK2}. An additional signature of this resonance is the ``mirroring'' behavior of the PCSs, a generic phenomenon that is ubiquitous in the following calculations \cite{CrossSecMirror,LiuHe}. The threshold behavior in the $5g$ channel is markedly different than in the $5f$ channel, and the slow climb above threshold was attributed to the repulsive polarization potential \cite{HanstorpK2}.

The dashed line in Fig. \ref{fig:5fzoom}b is a fit to the experimental data with the dipole TL modulated by a Shore profile describing the observed resonance \cite{SHORE,Note1}. This fit and that of Ref. \cite{HanstorpK2} are nearly indistinguishable. However, the latter fit yielded an atomic polarizability two orders of magnitude too small, suggesting that the agreement is fortuitous. In contrast, the successful fit to the dipole TL uses only an amplitude and the resonance profile as adjustable fit parameters, fixing $\lambda_\text{min} = 1.03$ (Table \ref{tab:polarizabilities}).

Further data is available in sodium. Fig. \ref{fig:5fzoom}c shows $5d$, $5f$, and $5g$ PCSs \cite{HanstorpNa}. The $5d$ and $5f$ PCSs were only measured near threshold \cite{RohlenThesis}. The calculations again reproduce the experimental observations. At lower energies the present calculations agree with Ref. \cite{LiuNa}, which did not study this higher energy range.  The $5d$ PCS rises sharply at threshold since both its polarization and dipole potentials are attractive. The $5g$ PCS rises slowly, consistent with repulsive polarization and dipole potentials. Several novel features that were not seen in potassium complicate the interpretation here. The $5f$ PCS rises initially, but then continues to climb slowly, apparently mixing aspects of both attractive and repulsive potentials. Furthermore, both its polarizabilities and dipole parameters are positive, leading to attractive polarization potentials but repulsive dipole potentials. Finally, the value for $\lambda_{min}$ taken from Tab. \ref{tab:polarizabilities} gives an unsatisfactory fit, whereas the fit shown in Fig. \ref{fig:5fzoom}c uses $\lambda_{min}=1.03$, as if only the $5f$ and $5g$ thresholds were degenerate. Again the polarization TL gives an $\alpha_p$ that is orders of magnitude too small \cite{HanstorpNa}.

\begin{figure}[t]
\begin{center}
\includegraphics[width =\columnwidth]{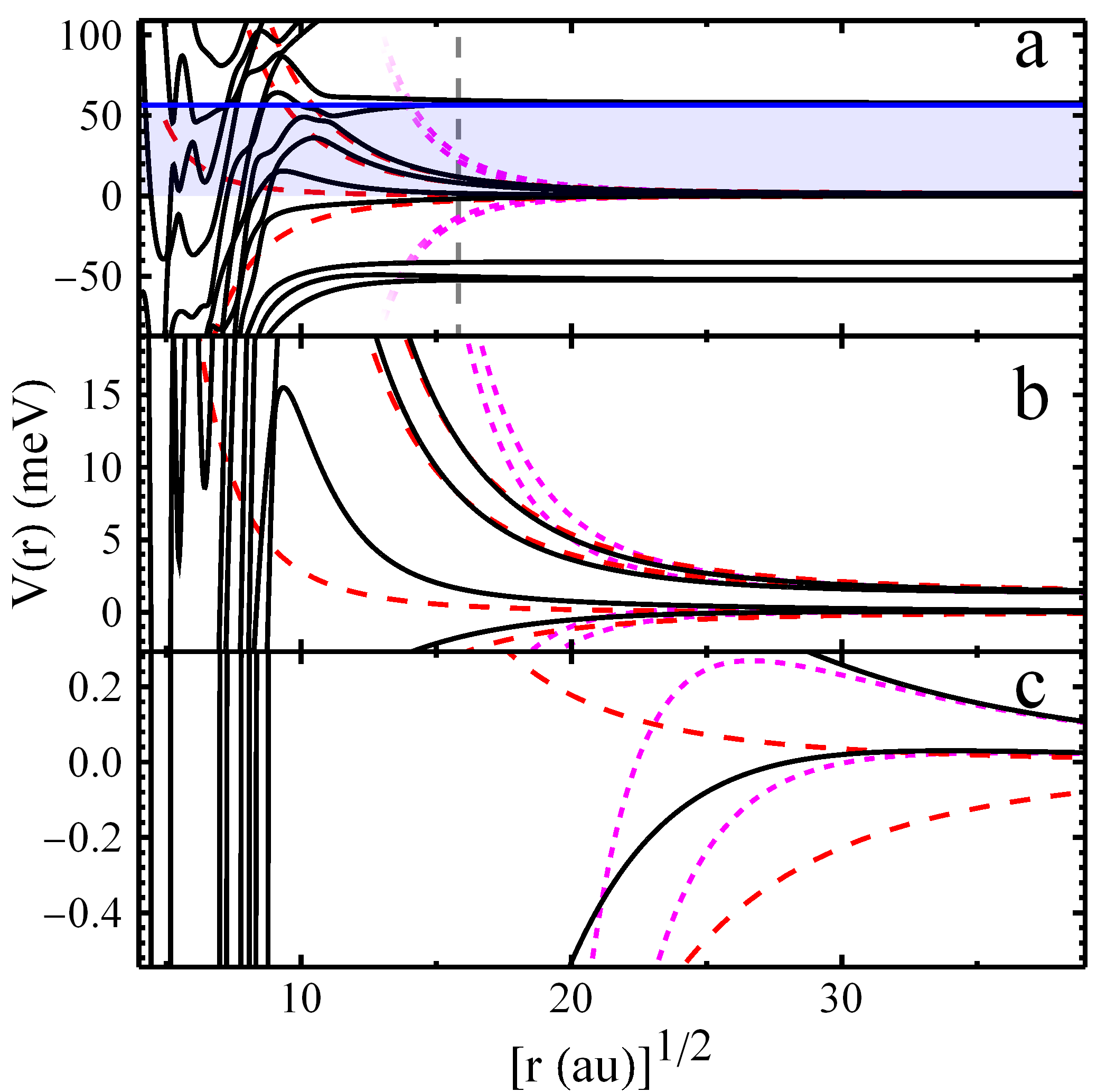}
\end{center}
\caption{\label{fig:PECsK1} Adiabatic potential energy curves (black,solid) for potassium, relative to the $5f$ threshold. a) The shaded blue region shows the energies measured in the experiment, and the vertical gray dashed line is at $r_0$.  The coarsely dashed red (finely dashed magenta) curves are the dipole (polarization) potentials.  b) and c) enlarge the region close to the $5f, 5g$ thresholds. }
\end{figure}

\begin{figure}[t]
\begin{center}
\includegraphics[width = \columnwidth]{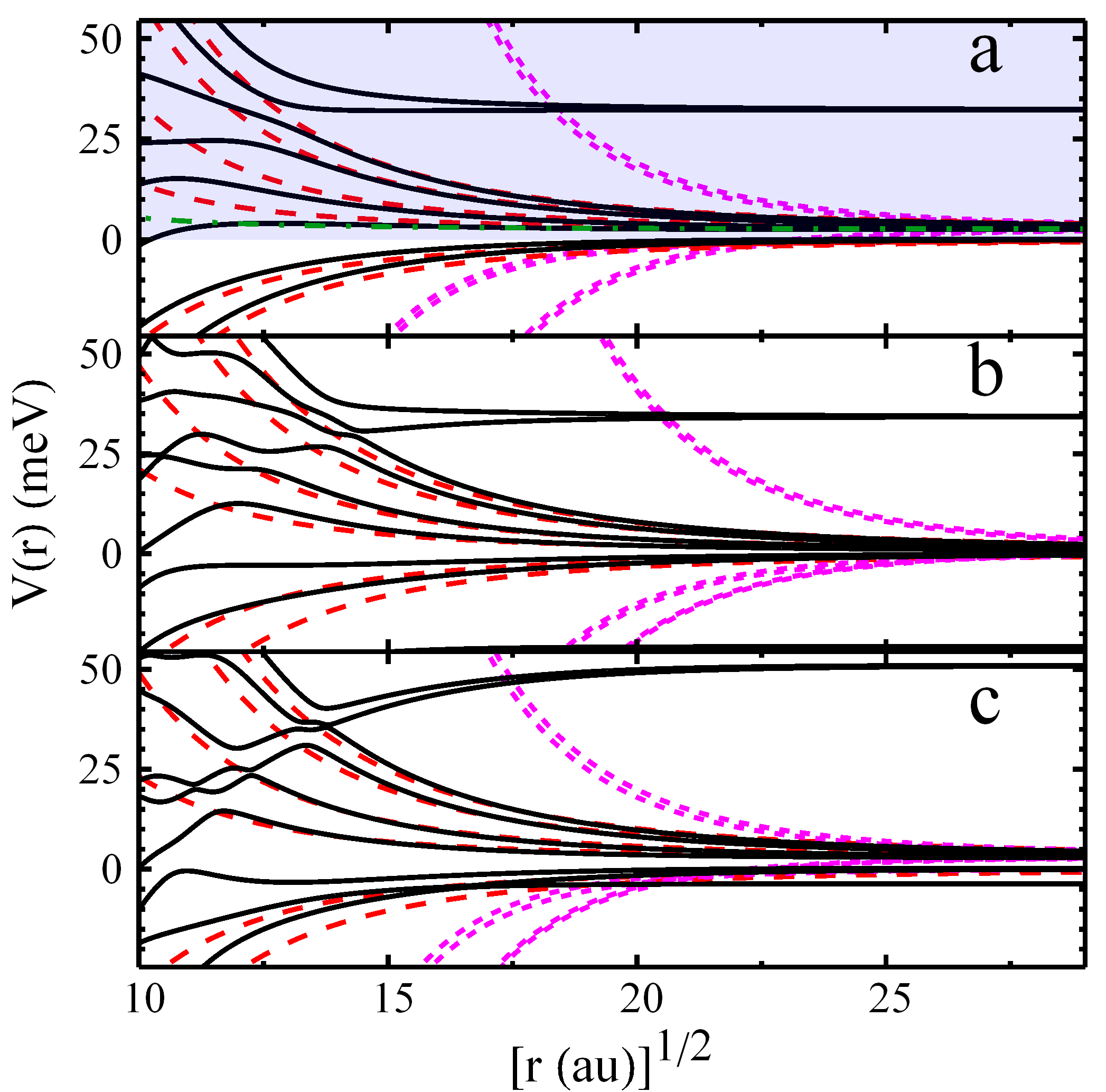}
\end{center}
\caption{\label{fig:PECsNa} Adiabatic potential energy curves (black) for (a) sodium, relative to the $5d$ threshold, (b) potassium, and (c) cesium, both relative to the $6f$ threshold. The region surveyed in the sodium experiment is shaded in blue. The red dashed (magenta) curves are the dipole (polarization) potentials. The green dot-dashed potential in panel (a) is a dipole potential for $\lambda = 1.03$, as discussed in the text.}
\end{figure}

Figs. \ref{fig:PECsK1} and \ref{fig:PECsNa}a show the adiabatic potential energy curves governing these processes, and a careful study of these curves resolves these complications.  First, these potentials justify the assumption of degenerate thresholds over the range of energies considered here: the $5f,5g$ splitting is indistinguishable on the scale of Fig. \ref{fig:PECsK1}a and relative to the range of energies explored experimentally. Additionally, the dipole potentials describe the adiabatic potentials far more accurately than the polarization potentials do, except at very low energy (panels b and c).   The potential curves for sodium shown in Fig. \ref{fig:PECsNa}a also lead to these conclusions, but also indicate why the $5f$ PCS is more challenging to match to the dipole TL and why a smaller $\lambda_\text{min}$ improves the fit. The $5d$ threshold is only approximately degenerate on this energy scale, so the potential curves are more poorly described by the dipole potentials than in potassium. Although these potentials are repulsive, and therefore the PCS should rise above threshold, this repulsion is very weak and the large energy separation between the $5d$ and $5f$ thresholds implies that the threshold behavior is not well described by pure dipole or polarization potentials. Excluding the $5d$ states from the degenerate subspace gives a dipole potential with $a_i = 2.09$ (dot-dashed green curve) which is qualitatively better and gives the satisfactory TL fit in Fig. \ref{fig:5fzoom}c.

\begin{figure}[t]
\begin{centering}
{\includegraphics[width=\columnwidth]{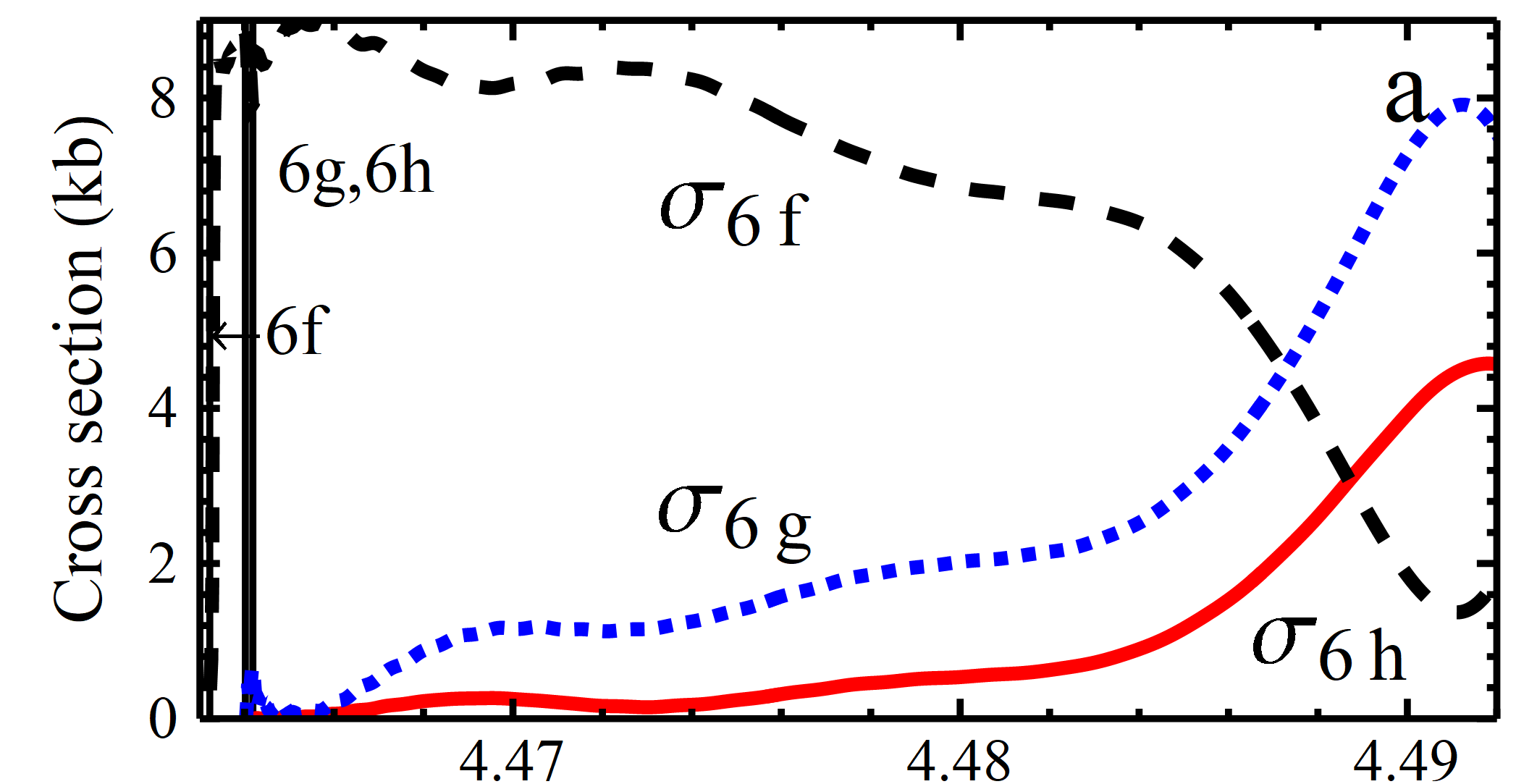}\\\includegraphics[width=\columnwidth]{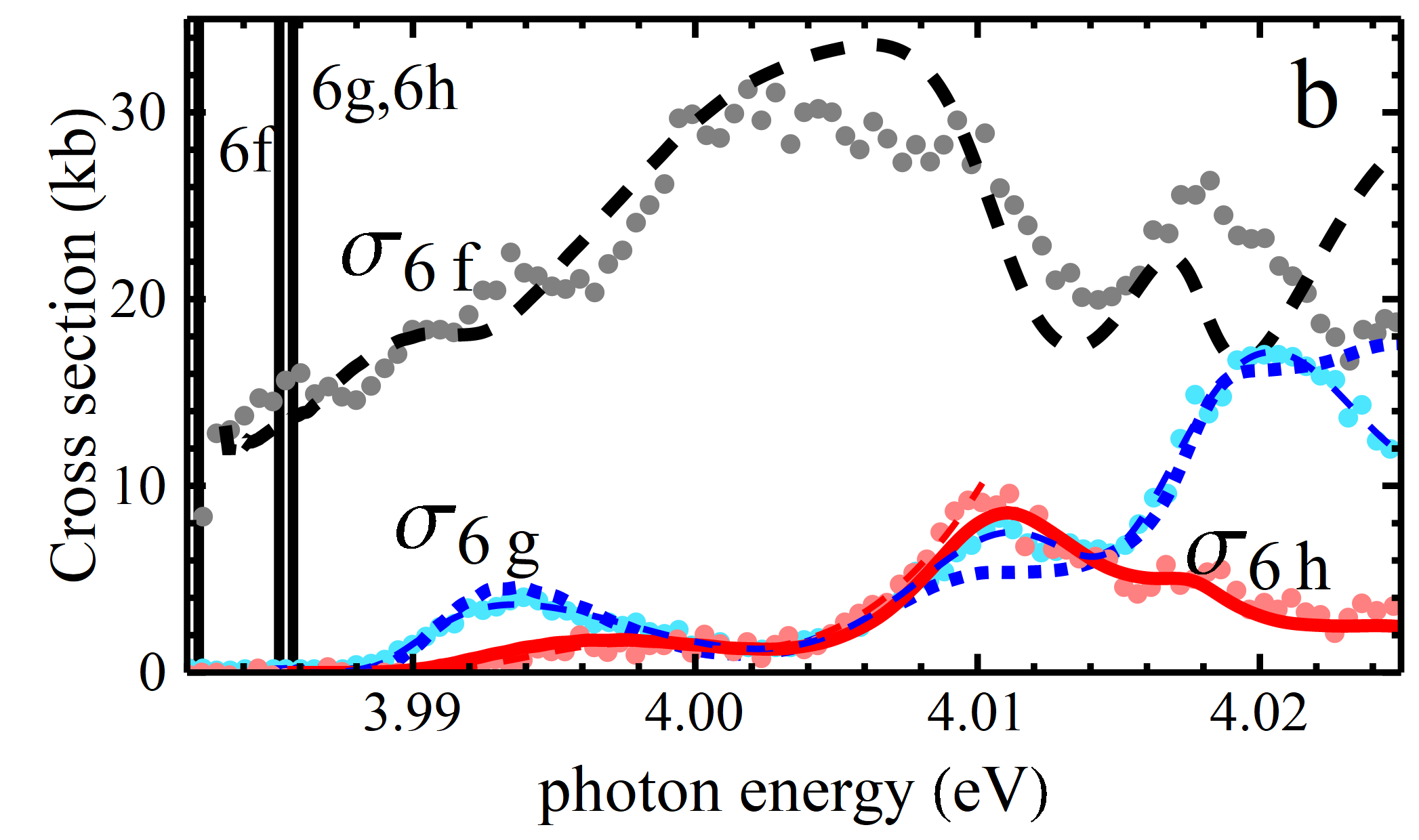}
}
\caption{ Observed \cite{HanstorpCsPRA} (round dots) and calculated PCSs for a) potassium and b) cesium. PCSs for the $6f$ (black, dashed line), $6g$ (blue, square dots), and $6h$ (red, solid) are shown. Thinner dashed curves show TL fits for the 6g and 6h PCSs; the $6h$ fit extends to 4.01 eV and is nearly indistinguishable from the calculation. }
\label{cesiumfigure}
\end{centering}
\end{figure}
The PCSs at the next atomic threshold elucidate further qualitative differences between dipole and polarization physics. These channels are more nearly degenerate, and inspection of Fig. \ref{fig:PECsNa} reveals that the potential curves are similar across atomic species due to the small quantum defects of these states. The polarization and dipole potentials again differ qualitatively: four attractive (two repulsive) polarization potentials and four repulsive (two attractive) dipole potentials arise (Tab. \ref{tab:polarizabilities}). Unlike in sodium the dipole potentials approximate the adiabatic potentials well, suggesting that the observed behavior of the PCSs here is unambiguously caused by dipole potentials in this degenerate subspace. 

Fig. \ref{cesiumfigure}a presents predicted PCSs in potassium. In contrast to the threshold behavior implied by the polarization potentials, but in accordance with the dipole potentials, only the $6f$ PCS begins at a finite value, while the $6g$ and $6h$ PCSs rise slowly. Fig. \ref{cesiumfigure}b shows the same channels in cesium along with measured results \cite{HanstorpCsPRA}. No previous calculations of these states exist. This calculation neglects relativistic spin-orbit effects, typically strong in heavy atoms like Cs, but these effects are reduced in channels with high $l$ and the calculations and observations are in good agreement. Again, these results agree only with the dipole potential predictions. They are successfully fitted to the dipole TL using $\lambda_\text{min} = 3.4$ (Tab. \ref{tab:polarizabilities}) and three Shore resonance profiles for the entire $6g$ PCS, while the $6h$ fit included only one resonance over a more limited range to better compare with the $5g$ channel fit in potassium \cite{Note1}. Fits using the polarization TL again significantly underestimated $\alpha_p$, and furthermore this TL is invalid for the $6g$ channel due to its positive polarizability. This conclusively shows that these experiments revealed repulsive dipole potentials, which will continue to control photodetachment at higher energy and angular momentum scales.

This Letter has elucidated the mechanism underlying the behavior of PCSs for photodetachment into channels with very high $l$. The calculations and experiment agree excellently, and the adiabatic potential energy curves are consistent with dipole potentials rather than with the polarization potentials typically dominant in non-hydrogenic atoms. Although the qualitative predictions of both potentials are consistent with observed $5d$, $5f$, and $5g$ cross sections, they are quantitatively much better described by the dipole TL. The $6f$, $6g$, and $6h$ PCSs of potassium and cesium provide clear confirmation of the formation of repulsive dipole potentials in this system, as the polarization potentials here are qualitatively wrong at energies above the tiny threshold splitting and lead to incorrect predictions in that range. This transition between two strikingly different power law potentials as the atomic core's excitation increases is, to our knowledge, the first observation of such an effect in an atomic system. Similar behavior has been observed previously in the photodetachment of molecular anions, which exhibit a transition from Wigner threshold behavior at very low energies to a non-Wigner threshold law at energies above the molecule's rotational splitting where the long-range potential of the electron becomes dipolar since the molecules possesses a dipole moment \cite{Engelking,LinebergerMol,HotopFabrikant,SadeghpourReview}. This transition is not limited to the single-photon detachment of alkali anions described here. Multiphoton photodetachment should exhibit this behavior and can access a variety of symmetries and parities, although the choice of intermediate state(s) could add additional complications. Other atomic species, particularly those in the copper and boron groups, ought to exhibit this same behavior since their anions are also effectively two-electron systems, although their more complex Rydberg structure, particularly in the Cu group due to the closed $d$ valence shell of its positive ion, could obscure this threshold behavior. 
\begin{acknowledgments}
 We are grateful to P. Giannakeas for his B-spline codes.  Preliminary work on this problem benefitted from discussions with the experimental team, especially J. Rohl\'{e}n and D. Hanstorp. Early stages of this work were begun during a KITP program, and we are grateful for the NSF support to fund that workshop. These calculations were performed using the computing cluster at the Purdue Rosen Center for Advanced Computing. This work was supported by the U.S. Department of Energy, Office of Science, under Award No. DE-SC0010545.
\end{acknowledgments}




%

\pagebreak
\widetext
\begin{center}
\textbf{\large Supplemental Information}
\end{center}
\setcounter{equation}{0}
\setcounter{figure}{0}
\setcounter{table}{0}
\setcounter{page}{1}
\makeatletter
\renewcommand{\theequation}{S\arabic{equation}}
\renewcommand{\thefigure}{S\arabic{figure}}
\renewcommand{\bibnumfmt}[1]{[S#1]}
\renewcommand{\citenumfont}[1]{S#1}

The parameters for the dipole threshold law fits described in the main text are listed here. As described in the main text, we used a Shore profile to describe the resonances \cite{SHORE}. For potassium, the $5g$ partial cross section was fitted to:
\begin{align*}
\sigma &= AE_e^{\lambda_\text{min}+1/2}\left(1 + \frac{a\epsilon +b}{1 + \epsilon^2}\right),\\\epsilon &= (E_\gamma - E_r)/(\Gamma_r/2),\,\,\lambda_\text{min} = 1.03,
\end{align*}
where $A$, $a$, $b$, $E_r$, and $\Gamma_r$ were all adjustable fit parameters. $E_\gamma$ here is the photon energy, $E_r$ is the resonance energy, and $\Gamma_r$ is the resonance width. $E_e$ is the electron energy,  $E_e = E_\gamma - E_t$, where $E_t$ is the threshold energy.  The fit displayed in the main text uses:
\begin{align*}
A &= 7.1163\times 10^{-16}\\
E_r &= 4.3199 \text{eV}\\
\Gamma_r &= 26.212\text{meV}\\
a &=-0.36174\\
b &=-0.89921.
\end{align*}
This resonance position and width are in excellent agreement with the fit performed in [17].
The $6h$ state of cesium was fitted similarly, with just one resonance, and with $\lambda_\text{min} = 3.4$. Here,
\begin{align*}
A &= 6.3530\times 10^{-8}\\
E_r &= 3.9960 \text{eV}\\
\Gamma_r &= 12.306\text{meV}\\
a &=-1.9601\\
b &=0.36942.
\end{align*}
Again, this resonance position and width match the fit performed in [30]. Extending the fit range would likely result in finding the second resonance above 4.01 eV, but at this point the presence of these resonances overcomplicates the fitting procedure here and obscures the close match of the dipole threshold law. 
The $6g$ channel could be fitted successfully over a larger region by including three resonances,
\begin{align*}
\sigma &= AE_e^{\lambda_\text{min}+1/2}\left(1 + \sum_{i=1}^3\frac{a_i\epsilon_i +b_i}{1 + \epsilon_i^2}\right),\\\epsilon_i &= (E_\gamma - E_{r_i})/(\Gamma_{r_i}/2),\,\,\lambda_\text{min} = 3.4.
\end{align*}
This summation over Shore profiles is applied regularly in scenarios where multiple resonances do not significantly overlap, e.g. as has been employed in Ref. \cite{HanstorpPRA}. These fit parameters were:
\begin{align*}
A &= 1.3634\times 10^{-7}\\
E_{r_1} &= 3.9896\text{eV}\\
\Gamma_{r_1} &= 6.4658 \text{meV}\\
a_1 &= -10.856\\
b_1 &= 22.570\\
E_{r_2} &= 4.01422\text{eV}\\
\Gamma_{r_2} &= 9.7006 \text{meV}\\
a_2 &=  -0.010077\\
b_2 &= -0.20098\\
E_{r_3} &= 4.01852\text{eV}\\
\Gamma_{r_3} &= 27.292 \text{meV}\\
a_3 &= -0.59251\\
b_3 &= 0.11258.
\end{align*}
These resonance widths and positions do not agree as well with those fitted in [30]: the first resonance is about $5$meV lower, but with a comparable width; the second resonance position and width are very comparable; the third resonance position is comparable but the width is much greater. The fits in [30] assumed only a linear function for the non-resonant cross section rather than any physically-motivated threshold law.

\end{document}